\documentstyle[12pt]{article}
\begin{document}

\ \\
\vskip 3.cm
\noindent
{\Large\bf Low-energy limit of the three-band model for electrons
 in a CuO$_{2}$ plane } \\
\vskip 1.cm

\noindent
{\large \bf V.I.Belinicher and A.L.Chernyshev } \\
\noindent
{\small \it Institute of Semiconductor Physics, 630090 Novosibirsk, Russia}\\
\vskip 1.cm

{\small The three-band model with the O-O direct hopping near to unit filling
 is considered. We present the general procedure of reduction of this model
to the low-energy limit. At unit filling the three-band model
in the charge-transfer limit is reduced to the Heisenberg model and we
calculate the superexchange constant. For the case of the small electron
doping the three-band model is reduced to the $t-J$ model and
we calculate electron hopping parameters at the nearest and next
neighbors. We derive the structure of corrections to the $t-J$ model
and calculate their magnitude. The values of the hopping parameters
for electron- and hole-doping differ approximately at 40 \%.}
\twocolumn
\vskip 1.cm
\noindent
{\bf 1. Introduction} \\
\vskip .1cm
Since the discovery of electron-doped superconductors \cite{Ta1} there
is growing interest in structure and properties of these systems
\cite{Th1,Lu1,J1}. The electron- and hole-doped high-tempera-
ture superconductors (such as Nd$_{2-x}$Ce$_{x}$CuO$_{4}$ and
 La$_{2-x}$Sr$_{x}$CuO$_{4}$) both have the CuO$_{2}$ planes with the
 same structure constant. In spite of the similar structure  the
experiments show the strong asymmetry in the properties of these compounds.
 Thus, the critical temperature for Nd$_{2-x}$Ce$_{x}$CuO$_{4}$ is only 22 K
that in contrast with T$_{c}$ = 40 K for  La$_{2-x}$Sr$_{x}$CuO$_{4}$
 \cite{Mu1}. Also, the doping concentration which destroy the AF order for
electron- and hole-doped systems are $\sim 15\%$ and $2\sim 3\%$
 respectively \cite{Th1,Lu1}.
To study of such asymmetrical phenomena might help to illuminate the
nature of superconductivity in these compounds.

Since the structure of the CuO$_{2}$ plane is the same for these
compounds, one can expect that well known three-band Hamiltonian \cite{Em1}
should be suitable for both of them.

In this paper we investigate one electron over unit filling
in a CuO$_{2}$ plane. We start from the Emery Hamiltonian \cite{Em1}
with the O-O direct hopping. On the basis of our general approach to the
low-energy reduction \cite{Be1}, we obtain the effective single-band
Hamiltonian. The equivalence of this single-band Hamiltonian and the
$t-J$ model is discussed and  the second-order corrections to the
$t-t'-J$ model are derived. The effective parameters of  the $t-t'-J$
model for the electron are calculated in the realistic region of Emery model's
parameters.
Effective hopping is less than in the case  of hole-doping.
Similar conclusion was obtained in the work by Zhang and Benneman \cite{ZB},
 but they used the perturbation theory over the parameters
 $t_{pd}/U_{d}$, $t_{pd}/(\epsilon_{p}-\epsilon_{d})$ which  does not
work as it was argued in our previous works \cite{Be1,Be2}.

In Sec.II we represent the three-band model in more suitable
terms for reduction to the single-band model. We use a three-step
procedure.  At the first step we introduce the symmetrical and
 antisymmetrical  oxygen operators. At the second step
we separate out the one-site
Hamiltonian and get its solution. At the third step we represent the primary
Hamiltonian in terms of the Hubbard operators which reflect the structure
 of the solution of the one-site problem.
In Sec. III we briefly consider the three-band model at unit filling.
Then, we apply the Schrieffer-Wolff transformation \cite{Bi1} for getting the
$J$-term  of the low-energy Hamiltonian.
In Sec. IV we get the effective electron hopping parameters and
correction to them. Also, we discuss the accuracy of the $t-t'-J$ models
for the electron-doped system. Section V presents our conclusions.
  In Appendix some details of our consideration are given.

\vskip 1.cm
\noindent
{\bf 2. Three-band Hamiltonian }  \\
\vskip .1cm
The three-band model with the direct O-O hopping is given by the following
Hamiltonian \cite{Em1}
\begin{eqnarray}
\label{1}
H&&=\epsilon_{d}\sum_{l,\alpha} d^{+}_{l\alpha}d_{l\alpha}+
\epsilon_{p}\sum_{m,\alpha}p^{+}_{m\alpha}p_{m\alpha}
\nonumber\\
&&+U_{d}\sum_{l}d^{+}_{l\uparrow}d_{l\uparrow}d^{+}_{l\downarrow}
d_{l\downarrow}+H' \ ,
\end{eqnarray}
where $d^{+}_{l\alpha}$ ($d_{l\alpha}$) creates (annihilates) the hole in a
 Cu $d_{x^{2}-y^{2}}$ state at site $l$, $p^{+}_{m\beta}$ ($p_{m\beta}$)
 creates (annihilates) the hole in a O $p_{x(y)}$ state at site $m$,
 $\epsilon_{d}$ and $\epsilon_{p}$ are the energies of Cu and O levels
respectively, $U_{d}$ is a intrasite Coulomb repulsion at the copper site.

The hybridization term $H'$ is given by
\begin{eqnarray}
\label{2}
H'&&=t\sum_{<lm>\alpha}(d^{+}_{l\alpha}p_{m\alpha}+ H.c.)
\nonumber\\
&&-t_{p}\sum_{<mm'>\alpha}(p^{+}_{m\alpha}p_{m'\alpha}+ H.c.) \ ,
\end{eqnarray}
 where $<lm>$ denote the nearest-neighbor Cu and O sites,
 $<mm'>$ denote the nearest-neighbor O sites.
In Eq. (\ref{2}) we follow to the sign convention of Ref.\cite{Em2,Fr1}.

As it was firstly noted by Zhang and Rice \cite{Zh1}, it is conveniently
to use the orthonormalized oxygen states on the four oxygens around a Cu site.
 In our previous work \cite{Be1} we have represented the Hamiltonian (\ref{1}),
(\ref{2}) in terms of orthogonal symmetrical and antisymmetrical oxygen
cluster states
\begin{eqnarray}
\label{2a}
q_{l}&&=\sum_{\bf k}(1+\gamma_{\bf k})^{-1/2}(\cos(k_{x}/2) p_{{\bf k}x}
\nonumber\\
&&+\cos(k_{y}/2) p_{{\bf k}y}) \exp(-i{\bf k}{\bf r}_{l}) \ ,
\nonumber\\
\tilde{q}_{l}&&=\sum_{\bf k}(1+\gamma_{\bf k})^{-1/2}(-\cos(k_{y}/2)
p_{{\bf k}x} \\
&&+\cos(k_{x}/2) p_{{\bf k}y}) \exp(-i{\bf k}{\bf r}_{l}) \ ,
\nonumber
\end{eqnarray}
where summation in Eq. (\ref{2a}) is produced over the Brillouin zone with
$\gamma_{{\bf k}}$=(1/2)(cos(k$_{x}$a)+\\
cos(k$_{y}$a)) and $p_{{\bf k} x,y}$
 are the Fourier image of $p_{m x,y}$
\begin{eqnarray}
\label{2b}
p_{{\bf k} x,y}=\sum_{m \in x,y} p_{m} \exp(-i{\bf k}{\bf r}_{m}) \ .
\end{eqnarray}
The physical reason for introduction of $q_{l}$ and $\tilde{q}_{l}$
states \cite{Zh1} is as follows.  The hole at the copper
can hope only at the
symmetrical combinations of the oxygen states. If we separate out  the
oxygen states which interact strongly with the copper states, we can
solve the problem of determination of the low-energy two-hole states
or vacuum states at the background of the one-hole states (spins).
 The states $q_{l}$, $\tilde{q}_{l}$ (\ref{2a}) are independent at different
sites and are ortonormalized. They are  very convenient for solving
this problem.

The original Hamiltonian (\ref{1}), (\ref{2}) in terms of $q_{l}$,
 $\tilde{q}_{l}$ takes the form
\begin{eqnarray}
\label{3}
H_{0}&&=\epsilon_{d}\sum_{l,\alpha} d^{+}_{l\alpha}d_{l\alpha}
+\epsilon_{p}\sum_{l,\alpha}(q^{+}_{l\alpha}q_{l\alpha}+
\tilde{q}^{+}_{l\alpha} \tilde{q}_{l\alpha})
\nonumber\\
&&+U_{d}\sum_{l}d^{+}_{l\uparrow}d_{l\uparrow}d^{+}_{l\downarrow}
d_{l\downarrow}
\nonumber\\
H'&&=2t\sum_{<ll'>\alpha}\lambda_{ll'}(d^{+}_{l\alpha}q_{l'\alpha}+ H.c.)
\\
&&-t_{p}\sum_{<ll'>\alpha}\{\mu_{ll'}(q^{+}_{l\alpha}q_{l'\alpha}-
\tilde{q}^{+}_{l\alpha} \tilde{q}_{l'\alpha})
\nonumber\\
&&+\nu_{ll'}(q^{+}_{l\alpha}\tilde{q}_{l'\alpha}+ H.c)\} \ ,
\nonumber
\end{eqnarray}
where the explicit form of the coefficients $\lambda_{ll'}$,
 $\mu_{ll'}$, $\nu_{ll'}$ can be obtained from Eqs. (\ref{1}), (\ref{2}),
 (\ref{2a})
\begin{eqnarray}
\label{4}
&&\{\lambda, \mu, \nu\}_{ll'} \equiv \{\lambda, \mu, \nu\}({\bf l}-{\bf l'})
\nonumber\\
 &&=\sum_{\bf k}\{\lambda, \mu, \nu\}_{\bf k} \exp(-i{\bf k}({\bf l-l'})) \ ,
\end{eqnarray}
with
\begin{eqnarray*}
\lambda_{\bf k}&&=(1+\gamma_{\bf k})^{1/2} \ ,\ \
\nonumber\\
\mu_{\bf k}&&=8\cos^{2}(k_{x}/2) \cos^{2}(k_{y}/2)
(1+\gamma_{\bf k})^{-1/2} \ , \\
\nu_{\bf k}&&=4\cos(k_{x}/2) \cos(k_{y}/2)(\cos^{2}(k_{x}/2)
\nonumber\\
&&- \cos^{2}(k_{y}/2)(1+\gamma_{\bf k})^{-1/2} \ .
\nonumber
\end{eqnarray*}
These coefficients decrease rapidly with increasing $|{\bf l}-{\bf l'}|$. The
values of $\lambda$, $\mu$ and $\nu$ for small $|{\bf l}-{\bf l'}|$ are
 given in Table I.
One can easily get that the $\nu_{00}\equiv 0$ and, therefore, the
mixing of $q$ and $\tilde{q}$ at the same sites is absent. The
transformed Hamiltonian (\ref{3}) is equivalent to the three-band
Hamiltonian (\ref{1}), (\ref{2}).

We divide the Hamiltonian (\ref{3}) into local and hopping parts
\begin{eqnarray}
\label{5}
H_{loc}&&=\epsilon_{d}\sum_{l,\alpha} d^{+}_{l\alpha}d_{l\alpha}+
(\epsilon_{p}-\mu_{0}t_{p})\sum_{l,\alpha}q^{+}_{l\alpha}q_{l\alpha}
\nonumber\\
&&+(\epsilon_{p}+\mu_{0}t_{p})\sum_{l,\alpha}\tilde{q}^{+}_{l\alpha}
 \tilde{q}_{l\alpha}
\nonumber\\
&&+U_{d}\sum_{l}d^{+}_{l\uparrow}d_{l\uparrow}
d^{+}_{l\downarrow}d_{l\downarrow}
\\
&&+2t\lambda_{0}\sum_{ll'\alpha}(d^{+}_{l\alpha}q_{l\alpha}+ H.c.) \ ,
\nonumber\\
H_{hop}&&=2t\sum_{ll'\alpha}\lambda_{ll'}(d^{+}_{l\alpha}q_{l'\alpha}+ H.c.)
\nonumber\\
&&-t_{p}\sum_{ll'\alpha}\{\mu_{ll'}(q^{+}_{l\alpha}q_{l'\alpha}-
\tilde{q}^{+}_{l\alpha} \tilde{q}_{l'\alpha})
\nonumber\\
&&+\nu_{ll'}(q^{+}_{l\alpha}\tilde{q}_{l'\alpha}+ H.c)\} \ ,
\nonumber
\end{eqnarray}
hereafter sum over $l,l'$ denotes $l\neq l'$.
 One can see that hybridization term in $H_{loc}$ (\ref{5}) includes only
the symmetric oxygen state in agreement with Zhang and Rice \cite{Zh1} .
 The direct O-O hopping, once taken into account, does not mix local states
with the opposite symmetry.

For the case of unit filling  there is one hole per unit cell. Therefore,
 one can introduce the set of one-hole cluster state
\begin{eqnarray}
\label{6}
&&|d_{\alpha}>\equiv d^{+}_{\alpha}|0> \  ,\
|q_{\alpha}>\equiv q^{+}_{\alpha}|0> \  ,  \
\nonumber\\
&&|\tilde{q}_{\alpha}>\equiv \tilde{q}^{+}_{\alpha}|0> \  ,
\end{eqnarray}
and rewrite $H_{loc}$ in this terms
\begin{eqnarray}
\label{7}
H^{1}_{loc}&&=\sum_{l,\alpha} \{ \epsilon_{d}
X^{dd}_{l\alpha }+(\epsilon_{p}-\mu_{0}t_{p}) X^{qq}_{l\alpha }
\nonumber\\
&&+(\epsilon_{p}+\mu_{0}t_{p}) X^{\tilde{q} \tilde{q}}_{l\alpha }
\\
&&+2t\lambda_{0}(X^{dq}_{l\alpha } + H.c.) \} \ ,
\nonumber
\end{eqnarray}
where
\begin{eqnarray}
\label{7a}
X^{ab}_{l\alpha }\equiv|a_{l\alpha}><b_{l\alpha}| \ \ ,
\end{eqnarray}
is the Hubbard operator at the site $l$, $\alpha=\pm\frac{1}{2}$ is
the spin projection. It is convenient to introduce the Hubbard operators
because they form the natural basis for description of one-site states. If
we introduce also the non-diagonal Hubbard  operators we can simply express
all operators in our Hamiltonian (\ref{5}) in their terms. Diagonalization of
$H^{1}_{loc}$ (\ref{7}) is performed for each site independently. After
diagonalization  $H^{1}_{loc}$ is given by
\begin{eqnarray}
\label{8}
H^{1}_{loc}&&=\sum_{l,\alpha} \left\{ \epsilon_{f}
X^{ff}_{l\alpha }+\epsilon_{g} X^{gg}_{l\alpha } \right.
\nonumber\\
&&\left. +(\epsilon_{p}+\mu_{0}t_{p}) X^{\tilde{q} \tilde{q}}_{l\alpha }
 \right\} \ ,
\end{eqnarray}
with
\begin{eqnarray}
\label{9}
&&|f_{\alpha}>=U |d_{\alpha}> - V |q_{\alpha}> \ ,
\nonumber\\
&&|g_{\alpha}>=V |d_{\alpha}> + U |q_{\alpha}> \ ,
\\
&&U=((R_{1}+\tilde{\Delta})/2R_{1})^{1/2} \ ,
\nonumber\\
&&V=((R_{1}-\tilde{\Delta})/2R_{1})^{1/2} \ ,
\nonumber
\end{eqnarray}
where $R_{1}=(\tilde{\Delta}^{2} + 4t^{2}\lambda_{0}^{2})^{1/2}$,
 $\tilde{\Delta}=(\Delta-\mu _{0}t_{p})/2$, and $\epsilon_{f,g}=-\Delta_{1}
 \mp R_{1}$, $\Delta_{1}=(\Delta+\mu _{0}t_{p})/2$, $\Delta=\epsilon_{p}-
\epsilon_{d}$. As discussed in Ref. \cite{Be1}, we assume that at unit
filling the groundstate of the CuO$_{2}$ plane is the low $|f\alpha >$ -
states on each cluster (the hole on the copper with admixture of
 symmetrical combination of the hole on the nearest four oxygens)
 with a virtual transitions at neighbors. Such transitions give the
superexchange interaction in the second order of perturbation theory.

The hopping part of the Hamiltonian (\ref{5}) contains transitions
between vacuum, one- and two-hole states at the different sites.  For a
detailed consideration of this subject see Ref. \cite{Be1}.

\vskip 1.cm
\noindent
{\bf 3. Superexchange interaction }   \\
\vskip .1cm
Here we consider only the terms of $H_{hop}$ that are relevant
to intersite interaction
\begin{eqnarray}
\label{10}
H_{hop}&&=\sum_{ll'\alpha \beta , y}F^{y,\alpha }_{ll',\beta}\left\{
X^{y,f\alpha}_{l}X^{0,f\beta}_{l'} \right.
\nonumber\\
&&\left.+ X^{f\beta,0}_{l'}X^{f\alpha,y}_{l}
\right\},
\end{eqnarray}
where $y$ is the set of two-hole states, which are five singlets and three
triplets (Ref. \cite{Be1}), $F^{y,\alpha }_{ll',\beta}$ are matrix elements
of the Hamiltonian for transition from $\{f_{l},f_{l'}\}$ to
$\{y_{l},0_{l'}\}$ states.

By applying the Schrieffer-Wolff transformation to (\ref{10}) (see Appendix
and \cite{Be1,Bi1}) one can get
\begin{eqnarray}
\label{11}
&&H_{J}=\sum_{ll'}(J_{ll'}{\bf S}_{l}{\bf S}_{l'}
+Y_{ll'}\hat{N}_{l}\hat{N}_{l'}) \ , \\
&&{\bf S}_{l}=(1/2){\bf \sigma}_{\alpha\beta}X^{\alpha \beta}_{l} \ , \ \
\hat{N}_{l}\equiv X^{\uparrow\uparrow}_{l}+X^{\downarrow\downarrow}_{l} \ ,
\nonumber
\end{eqnarray}
the expressions for the $J_{ll'}$ and $Y_{ll'}$ constant are given in
Appendix.

Thus, we established that at unit filling the groundstate of the CuO$_{2}$
plane is the system of local spins  which interact antiferromagnetically.
 Since $J_{ll'}$ decreases rapidly with increasing $|{\bf l}-{\bf l'}|$, we
hold only the nearest-neighbor term in Eq. (\ref{11}) and receive the
Heisenberg Hamiltonian. One can check that $Y_{<ll'>}\simeq -\frac{1}{4}
J_{<ll'>}$  and so
\begin{eqnarray}
\label{12}
&&H_{J}=J\sum_{<ll'>}({\bf S}_{l}{\bf S}_{l'}-\frac{1}{4}
\hat{N}_{l}\hat{N}_{l'}) \ ,
\nonumber\\
&&J\equiv J_{<ll'>} .
\end{eqnarray}
The second term in the Hamiltonian (\ref{12}) can be essential when the
system goes away from unit filling.

\vskip 1.cm
\noindent
{\bf 4. Electron-doped system } \\
\vskip .1cm
The doping of the CuO$_{2}$ plane by the electron is equivalent to removal
of a hole from one of the clusters. Hence, in the hole picture the electron
is 'zero' or vacuum state. If the hole (spin) is taken away from cluster, the
neighboring holes (spins) can hope to the empty site. Thus, the mechanism of
the movement of the electron is different from the movement of the added
hole. The electron moves as a 'hole-in-hole', whereas the hole moves
as a local singlet.

\vskip 1.cm
\noindent
{ \it 4.1. Hopping Hamiltonian }\\
\vskip .1cm
{}From expression for $H_{hop}$ (\ref{5}) with definitions (\ref{9})  and
(\ref{7a}) one can easily get the zero-order electron hopping Hamiltonian
\begin{eqnarray}
\label{13}
H_{t}=\sum_{ll'\alpha}T_{ll'}X^{0,f\alpha }_{l}X^{f\alpha,0}_{l'} \ ,
\end{eqnarray}
with
\begin{eqnarray}
\label{14}
T_{ll'}^{e}=-4t\lambda _{ll'} UV - t_{p}\mu _{ll'} V^{2} \  .
\end{eqnarray}
The electron hopping Hamiltonian (\ref{13}) coincides with the hole
 hopping Hamiltonian derived in \\ works \cite{Be1,Be2}. They
correspond
to the $t-t'-\ldots $  terms of the $t-t'-J$ Hamiltonian. But the
expression for the hole-singlet  hopping is more complicated \cite{Be1}
\begin{eqnarray}
\label{15}
T_{ll'}^{h}&&=2t\lambda _{ll'}(\sqrt{2}U_{1}U-W_{1}V)\times
\nonumber\\
&&\times (\sqrt{2}V_{1}V-W_{1}U)
\\
&&- t_{p}\mu _{ll'}
(\sqrt{2}V_{1}V-W_{1}U)^{2}/2  \ ,
\nonumber
\end{eqnarray}
where $U$ and $V$ were determined in Eq. (\ref{9}), $U_{1}$, $V_{1}$,
 $W_{1}$ are the components of eighenfunction of the local singlet.

Thus, we mapped the three-band model for one electron over unit filling
to the $t-t'-J$ model. The Hamiltonian (\ref{12}),(\ref{13}) in
terms of the Hubbard operators can be rewritten in a more usual form
\begin{eqnarray}
\label{15a}
H_{t-J}&&=t\sum_{<ll'>,\alpha}\tilde{c}^{+}_{l\alpha} \tilde{c}_{l'\alpha}
\nonumber\\
&&+J\sum_{<ll'>}({\bf S}_{l}{\bf S}_{l'}-\frac{1}{4}
\hat{n}_{l}\hat{n}_{l'}) \ ,
\\
\tilde{c}_{l\alpha}&& =c_{l\alpha}(1-\hat{n}_{l,-\alpha}) \ , \ \
\tilde{c}^{+}_{l\alpha}=(\tilde{c}_{l\alpha})^{+} \ , \ \
\nonumber\\
{\bf S}_{l}&&=(1/2)c_{l}^{+}{\bf \sigma}c_{l} \ , \ \ \
\hat{n}_{l}=(c_{l}^{+}\cdot c_{l}) \ .
\nonumber
\end{eqnarray}
Here $c_{l\alpha}^{+},c_{l\alpha}$  are the electron creation
 and annihilation operators.

\vskip 1.cm
\noindent
{ \it 4.2. Second-order corrections } \\
\vskip .1cm
With the help of Scrieffer-Wolff transformation  we shall obtain the
second-order corrections to the $t-t'-J$ Hamiltonian. The corrections
to the first hopping parameter  will be small and the $t-J$ model may be
valid. The correction to the other hopping parameters are not relatively
small and, therefore, models with the hopping at the neighbors farther
than nearest must be more complicated than the simple $t-t'-\ldots \ $  type.

First of all we shall obtain correction to the energy of the electron.
 Such  correction arises due to the virtual process of hopping of
the electron  at the neighbors with transition in excited state and back.
 Corresponding Hamiltonian is given in Appendix. The resulting
second-order correction is
\begin{eqnarray}
\label{16}
\delta H_{E}=-\sum_{ll'}M_{ll'}X^{00}_{l}\hat{N}_{l'}
\end{eqnarray}
with
\begin{eqnarray}
\label{17}
M_{ll'}&&=\frac{|T_{ll',f}^{g}|^{2}}{\epsilon_{g}- \epsilon_{f}}, \
\nonumber\\
T_{ll',f}^{g}&&=2t\lambda _{ll'} (U^{2}-V^{2})
\\
&&+ t_{p}\mu _{ll'} UV.
\nonumber
\end{eqnarray}

The Hamiltonian $\delta H_{E}$ (\ref{16}) coincides completely with
 $\delta H_{E}$ in Ref. \cite{Be1}, but the expression (\ref{17}) for the
 $M_{ll'}$ is sufficiently more simple than for the hole one.

Corrections to the direct hopping parameters can be obtained by
transformation of the Hamiltonian (\ref{10}). The result is
\begin{eqnarray}
\label{18}
\delta H_{t}&=\sum _{lnl'\alpha \beta ,y}T_{lnl'}^{y}\{x_{y}
X^{0,f\alpha}_{l}X^{f\alpha, 0}_{l'}\hat{N}_{n}
\nonumber \\
&+z_{y}X^{0,f\beta}_{l}X^{f\alpha,0}_{l'}
({\bf \sigma}_{ \alpha \beta}{\bf S}_{n})\} \ ,
\end{eqnarray}
where $y$ is the set of two-hole states, $x_{y}=1/2$ for singlets
and $x_{y}=3/4$ for triplets,$z_{y}=1$ for singlets and $z_{y}=-1/2$ for
triplets. The expressions for $T_{lnl'}^{y}$ are given in Appendix. As
it was shown , the corrections to the $t-t'-J$ Hamiltonian
have a complicated nature and depend on the filling ($\hat{N}$-term) and
spin state of the neighbors (${\bf S}$-term). The corrections
for the hole single-band model \cite{Be1} and the electron one are quite
different.

\vskip 1.cm
\noindent
{ \it 4.3. Quantitative results }  \\
\vskip .1cm
We have argued \cite{Be1} that our general approach to the low-energy
reduction of a three-band model is essentially more correct than the
other one, because we construct the perturbation theory over the
ratio of the effective hopping parameters between the different local states
 to the energy gap between them. This ratio for the case of CuO$_{2}$
 plane is of the order of 0.1 . All values of the effective single-band
Hamiltonian have been derived for arbitrary parameters of the three-band
model, without any assumption about a relationship among $t$, $\Delta$ and
$U_{d}$ . In our calculation we take $U_{d}=8.2$ eV, $t=1.4$ eV,
 $t_{p}=0.7$ eV,   $U_{p}=V_{pd}=0$ according to the different papers
 \cite{Pi1,Fl1,Fi1}. Also, we have taken experimental value of  $J=126$ meV
 \cite{Ma1} and have determined selfconsistently charge-transfer gap to be
$\Delta =5.07$ eV. In Table II we present the following effective parameters
 of one-band model: the hopping parameters to the first, second and third
neighbors; the contribution of the direct O-O hopping; corrections
 to hopping and energy ; value of the ratio $t_{1}/J$.

We  obtain for the electron hopping t$^{e}\equiv$ t$_{1}\approx 0.39$ eV
that in 1.36 times less than the hole hopping \cite{Be1}
 t$^{h}\approx 0.53$ eV . The role of the direct O-O hopping
for the electron-doped system is lower than for hole one.
 Actually, the admixture of the O-state for one hole on a cluster is small,
 unlike for the hole singlet, where added hole mainly on the oxygen.
 Thus, the O-O hopping contribution to the hopping at the
nearest-neighbors for electron (t$_{p1}$/t$_{1}$)$^{e}\approx20\%$ and for
hole (t$_{p1}$/t$_{1}$)$^{h}\approx36\%$.

As for the validity of $t-J$, $t-t'-J$ and other models to the
 electron-doped system, our conclusions are similar to the case of the
hole-doped system. Thus, if we hold only the hopping at the
nearest-neighbors, corrections to the $t-J$ model on a ferromagnetic
background is near to $\ \ 0.13\%$ for electrons and $\ \ 2.5\%$
 for holes, and $t-J$ model is valid. If we are interested in the
next-neighbors hopping, it is necessary to include the correction
  (\ref{18}) to the effective Hamiltonian. This correction has the
complicated structure and does not reduce to the simple direct hopping.

\vskip 1.cm
\noindent
{\bf 5. Conclusion } \\
\vskip .1cm
We have studied the electron doping of the CuO$_{2}$ plane in the framework
 of three-band model. By applying our general procedure of the low-energy
reduction to the electron-doped system, we conclude that the $t-J$ model
is valid as well as to the hole-doped system. We established some
asymmetrical properties. Effective hopping parameter for the electron
 is in 1.36 times less than the hole one. Corrections to the single-band
electron and hole $t-J$ models have the different magnitudes. Such
corrections are important for transitions at the next-nearest neighbors.

It is doubtful that the difference in the parameters of one-band
model may lead to the drastic difference in behavior
of hole-doped and electron-doped systems with doping. The \\ strong
asymmetrical dependence of the critical concentration for disappearance
of antiferromagnetizm is probably connected with the nature of
antiferromagnetic ordering in this systems \cite{Mat}. In the hole-doped
systems antiferromagnetizm has a quasi-two-dimensional character and
is associated with $Cu$ spins. In the electron-doped systems the
contribution of  $Nd$ or $Pr$ spins in the antiferromagnetic ordering
is essential and three-dimensional effects are more important.

\vskip 1.0cm
\noindent
{\bf  Acknowledgments} \\
\vskip 0.1cm
We would like  to thank O.P.Sushkov for the helpful discussion. This work
was supported by the Program  'Universities of Russia as the centers of
fundamental reseaches' and by the Counsel on Superconductivity
of Russian Academy of Sciences, Grant No. 90214.

\vskip 2.0cm
\noindent
{\bf Appendix: The second-order corrections to the hopping process }\\
\vskip .1cm
For obtaining such corrections we use the \\ Schrieffer-Wolff
transformation
\begin{eqnarray}
\label{a1}
H \Rightarrow \tilde{H}&&=\exp{(-S)} H \exp{(S)},
\nonumber \\
 S^{+}&&=-S .
\end{eqnarray}
The first order generator of transformation and the second-order
correction are
\begin{eqnarray}
\label{a2}
[H_{0},S]=-H', \ \ \ \delta H^{2}=(1/2)[H',S] \ ,
\end{eqnarray}
for this model $H_{0}\equiv H_{loc}$ (Eq.(\ref{5})). For the derivation
of $J$-term $H'\equiv H_{hop}$ (Eq.(\ref{10})).  Expressions for the
$J_{ll'}$ and $Y_{ll'}$ integrals are given by
\begin{eqnarray}
\label{a3}
J_{ll'}&&=\sum_{y}x_{y}\frac{|F_{ll'}^{y}|^{2}}{\epsilon_{y}+\epsilon _{0}
- 2\epsilon_{f}} \ , \
\nonumber\\
Y_{ll'}&&=-\sum_{y}z_{y}\frac{|F_{ll'}^{y}|^{2}}{\epsilon_{y}+\epsilon _{0}
- 2\epsilon_{f}} \ , \
\end{eqnarray}
where $y$ is the set of two-hole states at a cluster, $\epsilon _{y}$
 are there eighenenergies, $\epsilon_{0} \equiv 0$ is the energy of
vacuum state. Values $x_{y}=2$, $z_{y}=1/2$ for singlets and $x_{y}=-1$,
$z_{y}=3/4$ for triplets. Matrix elements  $F_{ll'}^{y}$ are obtained
in Ref. \cite{Be1}. One can check that the major contribution to $J$ and $Y$
 is originated from the matrix element of the transition to the lowest
 singlet and so $Y_{<ll'>}\simeq \frac{1}{4} J_{<ll'>}$.

The terms of $H_{hop}$ (\ref{5}), which lead to the energy correction, are
given by
\begin{eqnarray}
\label{a4}
H'_{E}&&=\sum_{ll'\alpha}F^{g}_{ll',f}\left\{
X^{0,f\alpha}_{l}X^{g\alpha,0}_{l'}\right.
\nonumber\\
&&\left.+ X^{0,g\alpha}_{l'}X^{f\alpha,0}_{l}
\right\} \ .
\end{eqnarray}
Corresponding generator of the transformation is
\begin{eqnarray}
\label{a5}
S_{E}&&=-\sum_{ll'\alpha}\frac{F^{g}_{ll',f}}{\epsilon_{g}
-\epsilon _{f}}\left\{
X^{0,f\alpha}_{l}X^{g\alpha ,0}_{l'} \right.
\nonumber\\
&&\left. - X^{0,g\alpha }_{l'}X^{f\alpha,0}_{l}
\right\} \ ,
\end{eqnarray}
where $|f_{\alpha}>$ and $|g_{\alpha}>$ are the one-hole states (\ref{9})
 with their energies $\epsilon _{f}$ and $\epsilon _{g}$.
Thus, the second-order correction to the energy has the form (\ref{16}).

Corrections to the direct hopping (\ref{18}) can be obtained
 from the transformation
of the \\ Hamiltonian (\ref{10}). Magnitudes $T^{y}_{lnl'}$ are given by
\begin{eqnarray}
\label{a6}
T_{lnl'}^{y}=\frac{F_{ln}^{y}F_{nl'}^{y}}{\epsilon_{y}+\epsilon _{0}
- 2\epsilon_{f}} \ .
\end{eqnarray}
The expressions of  $F_{ln}^{y}$ were obtained  analytically and
computed for definite values of parameters of the three-band model in
Ref. \cite{Be1}.

\onecolumn
\noindent
Table I.  \\ {\small The values of the coefficients
$\lambda ({\bf l}-{\bf l'})$,
 $\mu ({\bf l}-{\bf l'})$, $\nu ({\bf l}-{\bf l'})$ as  functions of
 $({\bf l}-{\bf l'})=n{\bf x}+m{\bf y}$ }.
\vskip 0.5 cm
\begin{center}
\begin{tabular}{cccc}
\hline \\
 n,m  & $\lambda _{n,m}=\lambda _{m,n}$ & $\mu _{n,m}=\mu _{m,n}$ &
$\nu_{n,m}=-\nu _{m,n}$ \\  \\ \hline \\
 0 , 0  & 0.9581 & 1.4567 & 0.0  \\
 1 , 0  & 0.1401 & 0.5497 & 0.2678   \\
 1 , 1  & -0.0235 & 0.2483 & 0.0      \\
 2 , 0  & -0.0137 & -0.1245 & 0.0812    \\
 2 , 1  & 0.0069 & -0.0322 & 0.0609      \\
 2 , 2  & 0.0035 & 0.0231 & 0.0  \\ \\ \hline
\end{tabular}
\end{center}
\vskip 0.5 cm

\noindent
Table II. \\ {\small The first three hopping parameters, the contributions
of the direct O-O hopping to them, second-order corrections to them
on a ferromagnetic background, second-order corrections to the energy,
and the ratio $t_{1}/J$.}
\vskip 0.5 cm
\begin{center}
\begin{tabular}{cccc}
\hline \\
 neighbor & direct hopping & direct O-O hopping &  Corrections  \\
 number, n & $t_{n}^{eff}$, (eV) & $t_{pn}^{eff}$, (eV) &
$\delta t_{n}^{eff}$, ($\%$) \\ \\ \hline \\
 1  & -0.3896 & -0.0765 & 0.125 \\
 2  & 0.0180 & -0.0345 & 122.5  \\
 3  & 0.0480 & 0.0173 & 24.2  \\ \\  \hline \\
  Correction to &  & Ratio &  \\
  the energy (eV) & -0.0931 & $|t_{1}/J|$ & 3.081 \\ \\ \hline
\end{tabular}
\end{center}
\vskip 0.5 cm

\newpage
\vspace{2.cm}
\begin{thebibliography}{99}
{\small
\bibitem{Ta1} H.Takagi, S.Uchida and Y.Tokura, Phys. Rev. Lett.
  62 (1989) 1197.

\bibitem{Th1} T.R.Thurston et al., Phys. Rev. Lett.  65 (1990) 263.

\bibitem{Lu1} G.M.Luke et al., Phys. Rev. B 42 (1990) 7981.

\bibitem{J1} Jian Ping Lu and Qimiao Si, Phys. Rev. B 42  (1990) 950.

\bibitem{Mu1} J.G.Bendors and K.A.M\"{u}ller, Z. Phys. B 64
 (1986) 189.

\bibitem{Em1} V.J.Emery, Phys. Rev. Lett. 58 (1987) 2794;
C.M.Varma, S.Schmitt-Rink and E.Abrahams, Solid State
Commun. 62 (1987) 681.

\bibitem{Be1} V.I.Belinicher and A.L.Chernyshev, (to be published).

\bibitem{ZB} Weiyi Zhang and K.Bennemann, Phys. Rev. B 45
(1992) 12487.

\bibitem{Be2} V.I.Belinicher and A.L.Chernyshev, Phys. Rev. B 46
(1992) ....

\bibitem{Bi1} G.L.Bir and G.E.Pikus, Symmetry and deformational effects
in Semiconductors (Wiley, New York, 1974).

\bibitem{Em2} V.J.Emery and G.Reiter, Phys.Rev. B 38
 (1988) 4547.

\bibitem{Fr1} P.M.Frenkel and R.J.Gooding, B.I.Shraiman, and
E.D. Siggia, Phys.Rev. B 41 (1990) 350.

\bibitem{Zh1} F.C.Zhang and T.M.Rice, Phys. Rev. B 37
 (1988) 3759.

\bibitem{Pi1} W.E.Pickett, Rev.Mod.Phys. 61 (1989) 433.

\bibitem{Fl1} V.V.Flambaum and O.P.Sushkov, Physica C 175 (1991) 347.

\bibitem{Fi1} J.Fink at al., Physica C 185-189 (1991) 45.

\bibitem{Ma1} S.Maekawa, Y.Ohita, and T.Tohyama, Physica C 185-189
 (1991) 168.

\bibitem{Mat} M.Matsuda, Y.Endoh, and K.Yamada, H.Kojima and I.Tanaka,
 R.J.Birgeneau and M.A.Kastner, G.Shirane, Phys. Rev. B 45 (1992) 12548.
 }
\end{thebibliography}
\end{document}